\documentclass[twoside,slac_one]{revtex4}
\usepackage{graphicx}
\usepackage{fancyhdr}
\usepackage{amsmath} 
\usepackage{bm}
\usepackage{amsxtra}
\usepackage{amssymb}
\usepackage{amsthm}
\usepackage{latexsym}
\usepackage{lscape}

\begin{document}

\title{The DarkSide Program at LNGS}

%

\author{A. Wright,}
\affiliation{Department of Physics, Princeton University, Princeton, NJ, USA,}
\author{for the DarkSide Collaboration.}

\begin{abstract}
DarkSide is a direct detection dark matter program based on two phase
time projection chambers with depleted argon targets. The DarkSide
detectors are designed, using novel low background techniques and
active shielding, to be capable of demonstrating {\it in situ} a very
low level of residual background. This means that each detector
in the DarkSide program should have the ability to make a convincing
claim of dark matter detection based on the observation of a few
nuclear recoil events. The collaboration is currently operating a
10\,kg prototype detector, DarkSide-10, in Laboratori Nazionali del Gran
Sasso (``LNGS''), Italy, while the
first physics detector in the program, DarkSide-50, is expected to be
deployed at LNGS at the end of 2012.
\end{abstract}

\maketitle

\thispagestyle{fancy}


\section{Introduction}

Cosmological and astrophysical evidence now overwhelmingly supports
the existence of ``dark matter,'' an as yet unknown form of matter, with
a total energy density roughly five times that of baryonic
matter. Direct detection dark matter programs attempt to
detect one well motivated dark matter candidate, Weakly
Interacting Massive Particles (WIMPs). If WIMPs exist, they should,
very occasionally, interact with an atomic nucleus causing the nucleus
to recoil with a small, but detectable, amount of kinetic energy.

DarkSide will attempt to detect WIMP-induced nuclear recoils using
two-phase depleted argon time projection chambers (TPCs). Argon TPCs,
which are described in detail in \cite{warp} and \cite{luca}, measure,
by optical means, both the energy deposited and the ionization charge
produced by particle interactions in their liquid active volumes. 

The largest challenge in searching for dark matter is the
suppression of the rate of background events to below the very low WIMP
interaction rates (a few events per ton-year) to which current dark
matter experiments are sensitive. One reason that argon is a promising
medium for dark matter searches is that it provides the ability, using
pulse shape discrimination (``PSD'') based on the
time profile of the the primary scintillation signal, to reject
electron recoil background events to levels in excess of 
$10^{-8}$ (\cite{boulay, DEAP, DEAP1, warp}). In two-phase operation, additional
discrimination, at the level of $10^{-2}$ to $10^{-3}$, can be achieved
using the charge-to-energy ratio of each event (\cite{warp, luca}).

DarkSide will build on this electron recoil discrimination to develop
a series of dark matter detectors with extraordinarily low, and very
well understood, rates of background events. Like all dark matter
experiments, the DarkSide detectors will be constructed from materials
with the lowest possible levels of intrinsic radioactivity and will be
assembled with great care to prevent them from being radioactively
contaminated in any way. Two novel low
background techniques that will be used in DarkSide are argon
naturally depleted in radioactive $^{39}$Ar, and ultra-low background
photodetectors. These are discussed in Sections \ref{sec:ar} and \ref{sec:pmt}.

\begin{figure}[htb]
\centering
\begin{minipage}[c]{7cm}
  \includegraphics[height=.22\textheight]{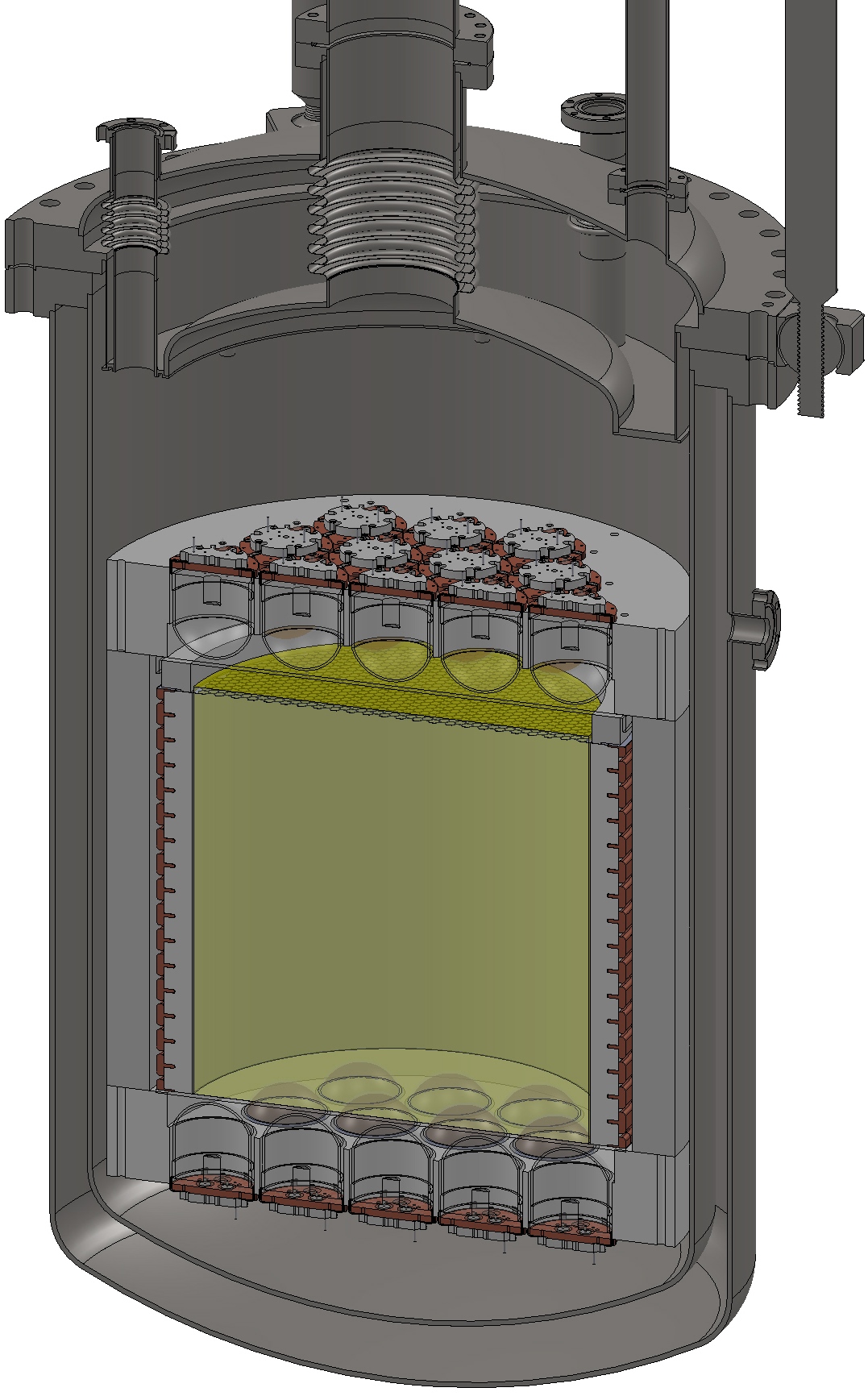}
\end{minipage}
\begin{minipage}[c]{7cm}
  \includegraphics[height=.25\textheight]{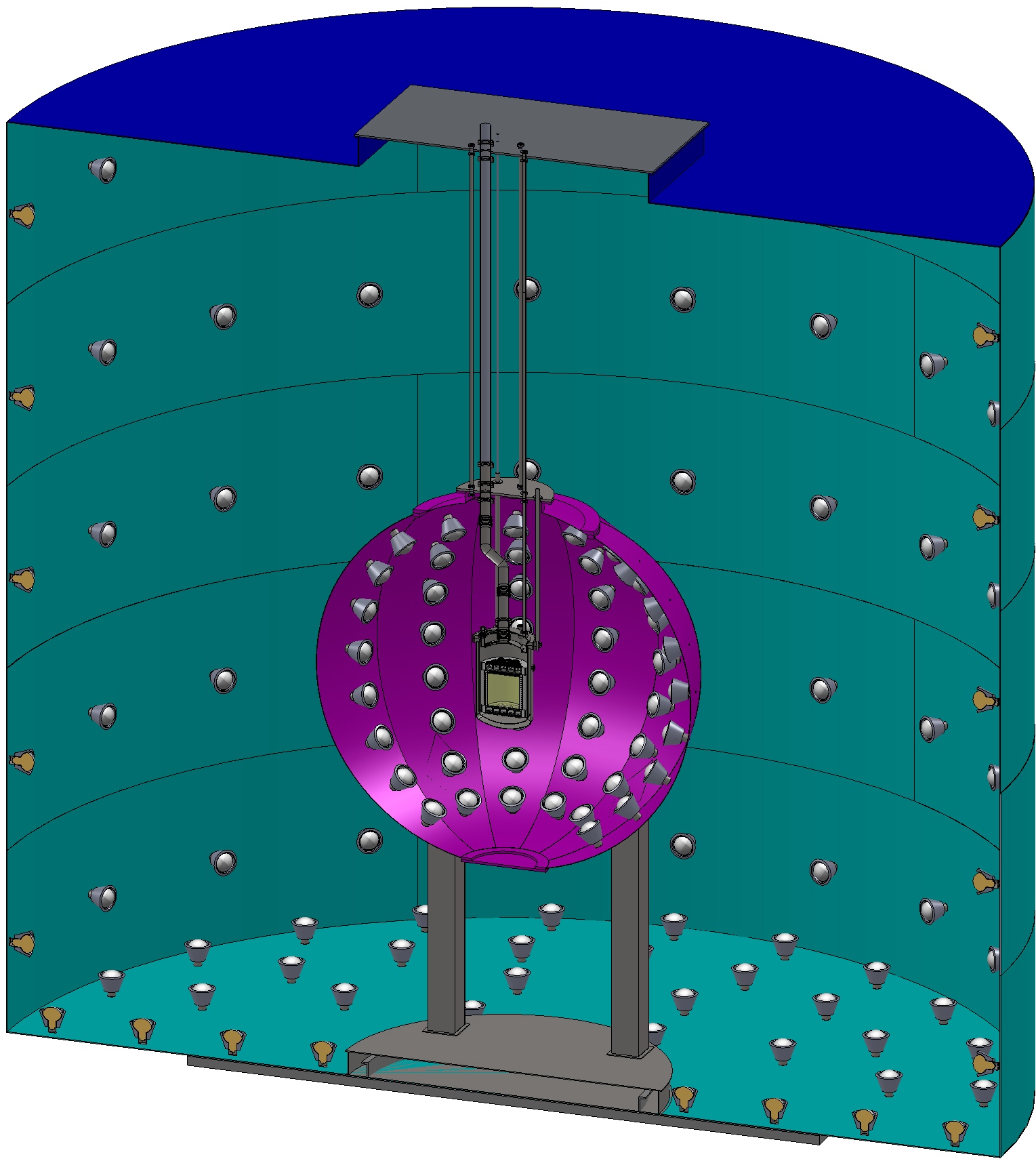}
\end{minipage}
\caption{Left: A conceptual drawing of the DarkSide-50 TPC in its dewar. The liquid argon
  volume is surrounded by an optical reflector and viewed by two arrays
of photodetectors. All surfaces facing the active volume are coated
with wavelength shifter. Right: The DarkSide-50
detector deployed within the active vetoes. The 4\,m diameter neutron
veto, filled with boron-loaded liquid scintillator, will be deployed
within the existing 11\,m diameter, 10\,m high water tank of the
Borexino CTF. The CTF tank will be instrumented to provide a water
Cerenkov muon veto.}
\label{fig:DS}
\end{figure}

Another important feature of the DarkSide design is the use of active
background suppression in place of passive shielding. As shown in Figure \ref{fig:DS}, DarkSide will be deployed within a
liquid scintillator based active neutron veto, which will in turn be
deployed within a water Cerenkov muon veto. The use of active
suppression will not only further reduce the already low background level,
but will also give DarkSide the capability of measuring the rates of the
different classes of background {\it in situ}. The DarkSide active background
suppression strategy is described in Section \ref{sec:bg}.

The DarkSide program will follow a staged approach. The first physics
detector in the program will be DarkSide-50, with a 50\,kg active
mass and a WIMP sensitivity of about 10$^{-45}$cm$^2$. DarkSide-50 is
currently under construction, with deployment planned for late 2012. Where possible, the DarkSide-50
infrastructure is being constructed such that a direct upgrade to a 5\,tonne
detector is possible. The collaboration is currently operating a
10\,kg prototype, DarkSide-10, at LNGS. DarkSide-10 is described in
Section \ref{sec:ds10}, while the future DarkSide program is outlined
in Section \ref{sec:dsFut}

\section{Depleted Argon}
\label{sec:ar}

Atmospheric argon contains $^{39}$Ar, which is a $\beta$ emitter with
an endpoint energy of 565\,keV, at the level of
$\sim$1\,Bq/kg~(\cite{warp39ar, loosli}). This
provides the dominant source of electron recoil backgrounds in
argon-based dark matter detectors. Although the $^{39}$Ar background
can be suppressed to levels acceptable for the operation of
tonne-scale detectors, it does impact the sensitivity of these
experiments by increasing their analysis thresholds\footnote{PSD
  efficiency is strongly dependent on photon statistics, so higher
  $^{39}$Ar rates may mean that
higher analysis thresholds are necessary to ensure sufficient PSD power to
suppress the $^{39}$Ar background.} and can make scaling to multi-tonne
experiments more difficult.

As the $^{39}$Ar halflife is only 269\,yr, the $^{39}$Ar activity in
the  atmosphere is maintained by cosmogenic production. Argon from
underground sources, which is shielded from cosmogenic activity, was demonstrated in~\cite{acosta}
to be depleted in $^{39}$Ar by at least a factor of 25 relative to
atmospheric argon. One such source is the Kinder Morgan Doe Canyon
Complex in Cortez, Colorado, which extracts CO$_2$ from natural underground
reservoirs. This gas contains $\sim$600\,ppm of argon. Since February 2010,
the DarkSide collaboration has operated an argon extraction facility
at the Kinder Morgan site~(\cite{ar_extract}). To date, about 46\,kg of
argon has been extracted.

\begin{figure}[htb]
\centering
\begin{minipage}[c]{7cm}
  \includegraphics[height=.22\textheight]{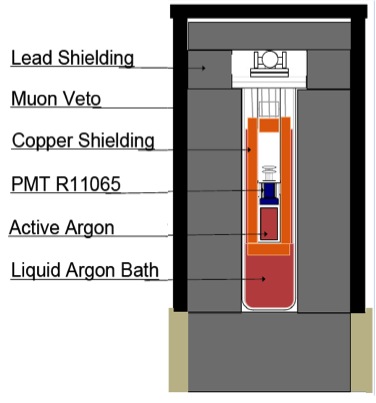}
\end{minipage}
\begin{minipage}[c]{7cm}
  \includegraphics[height=.25\textheight]{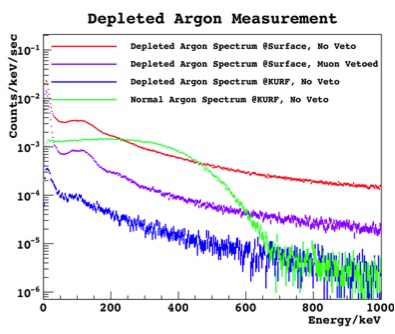}
\end{minipage}
\caption{Left: Schematic diagram of the ``Low Background Detector.'' Right:
  The depleted argon spectra obtained in various detector
  configurations. In the measurement at KURF, the total event rate in
  300-400~keV is $\sim$0.002\,Hz, about 2\% of the rate expected from
  $^{39}$Ar in atmospheric argon. Data taken with atmospheric argon is
  shown for comparison (green) - in this data the $^{39}$Ar spectrum is clearly visible.}
\label{fig:Ar}
\end{figure}

In order to study the residual $^{39}$Ar content in the Doe Canyon
argon, a dedicated low background detector has been
constructed. As shown in Figure \ref{fig:Ar}, the detector consists of
approximately 0.5\,kg of liquid argon coupled to a single 3''
cryogenic photomultiplier tube. The active volume is surrounded by 2''
of copper shielding and is immersed in a bath of commercial liquid argon,
which is allowed to evaporate to provide cooling. Eight inches of lead
shielding and a
plastic scintillator-based muon veto surround the argon bath.

The detector was first operated aboveground at Princeton
University, and a lower limit of
50 was placed on the depletion factor of $^{39}$Ar in the underground
argon relative to atmospheric argon. This limit was determined by fitting the
depleted argon energy spectrum (purple curve in Figure \ref{fig:Ar}) to
a smoothly decaying background shape and the expected $^{39}$Ar spectrum. As
the measurements on surface were
limited by non-vetoed cosmogenic background, the detector was moved
underground to the KURF laboratory in Kimballton, Va. (at a depth of 1400 meters of
water equivalent) in
the spring of 2011. As shown in Figure \ref{fig:Ar}, the background rate
at KURF was approximately 10 times lower than the background rate at Princeton. This lower
background allowed a lower limit of 50 to be placed on the $^{39}$Ar
depletion factor using direct counting alone; spectral fits, similar
to those used in the analysis of the Princeton data, are in progress,
and are expected to result in a significantly better limit.

\section{Low Background Photodetectors}
\label{sec:pmt}

Nuclear recoils produced by energetic neutrons which scatter only once
in the active volume form a background which is, on an event-by-event
basis, indistinguishable from dark matter interactions. Neutrons
capable of producing these recoil backgrounds are created by
radiogenic processes in the detector material; in detectors made from
clean materials, the dominant source of radiogenic neutrons is typically the
photodetectors.

With this limitation in mind, the UCLA DarkSide group has worked with Hamamatsu to
develop the Quartz Photon Intensifying Device, or ``QUPID.'' The QUPID
is a hybrid photomultiplier tube in which the photoelectrons are
  accelerated directly onto an avalanche photo-diode. As QUPIDs are
  constructed almost entirely of fused silica, they have the potential
  to be extremely ``clean'' from the point of view of radioactive
  contamination. QUPID
  development is ongoing, with extensive testing of production version
  QUPIDs currently underway at UCLA. QUPID design, development, and
  testing is described further in \cite{qupids}.

DarkSide has acquired a set of
Hamamatsu R11065 photomultiplier tubes for use while QUPID development is finalized. These 3'' metal bulb PMTs offer low
backgrounds ($\sim$60 mBq/PMT $\gamma$ activity and a neutron
production of $\sim$3 n/PMT/yr) coupled with high quantum efficiency ($\sim$35\% at
420\,nm) cryogenic operation. They are currently being used in the
DarkSide-10 prototype, and will be used in the initial deployment of DarkSide-50.

\section{Background Measurement and Suppression}
\label{sec:bg}

In the search for very rare events like WIMP interactions, the
credibility of a detection claim depends critically on
how well the background levels in the experiment producing
the claim are understood. At
current experimental sensitivities, it would be quite difficult to
make a convincing claim of dark matter detection after the observation
of a few recoil-like events if that claim were supported only by
background estimates based on Monte Carlo
calculations and ex-situ radioassays of representative material
samples. This is because, for the very clean materials used in the
construction of modern dark matter detectors, slight changes in the manufacturing
or handling of the material can cause the background rate in the material used in the
construction of the detector to differ critically from the activity in
the assayed samples, even under the most carefully controlled conditions.

The approach chosen by the DarkSide collaboration is to not only have the
lowest levels of background achievable, but also to incorporate powerful active
discrimination against each of the important classes of background
expected in the experiment. The active discrimination
has the potential to significantly suppress the backgrounds in
question, and also to provide
direct {\it in situ} measurements of the rate of each type of
background. This direct knowledge of the backgrounds, when combined with
calibration measurements demonstrating the effectiveness of the
active suppression techniques, allows the number of expected un-suppressed
background events to be convincingly determined at the
fraction-of-an-event level. Such a comprehensive understanding of the
backgrounds in an experiment would not only provide valuable guidance in the development of
next-generation experiments, but would also greatly increase the
credibility of a potential dark matter detection claim based on the observation
of a few events.

To suppress background events induced by radiogenic
neutrons, the DarkSide detectors will be deployed within a highly
efficient neutron veto made from boron loaded liquid scintillator. As
described in~\cite{veto}, Monte Carlo simulations suggest that the
neutron veto should achieve better than 99.5\% efficiency in rejecting
these background events. The use of boron-loaded liquid scintillator
significantly reduces the neutron capture time relative to pure
liquid scintillator, meaning that shorter veto windows are needed to
achieve high veto efficiency. This in turn means that the veto can be
simply constructed, with 8'' PMTs deployed directly in the veto
scintillator, without the resulting few hundred Hz trigger rate in the
veto inducing unacceptably high dead-time in the dark matter
detector. As shown in Figure \ref{fig:DS}, we plan to deploy DarkSide-50
within a 4\,m diameter neutron veto; this is large enough to allow the same
veto to be used for a future 5\,T detector.

Nuclear recoils induced by high energy cosmogenic neutrons will be suppressed by a water Cerenkov muon
veto which will be created by re-using the 10\,m high $\times$ 11\,m
diameter water tank
initially constructed for the Borexino Counting Test
Facility~(\cite{CTF}). The CTF tank will be retrofitted with the
addition of an optical reflector and a re-configuration of the existing
PMTs to optimize its detection efficiency for muons and other
cosmogenic shower particles. Conservative estimates show that the
combination of the water Cerenkov muon veto and the neutron veto will
suppress the cosmogenic background rate by at least a factor of 1000. 

Surface background events, in which $\alpha$ decays
from radon progeny on the inner surfaces of  the detector produce
nuclear recoil daughters that enter the active volume, are another
important class of background. They can be suppressed by fiducialization, with an attendant
loss in fiducial volume. Direct scintillation light production
by the passage of the alpha particle through the wavelength shifting
layer~(\cite{wls_scint}) will change the apparent energy and pulse shape
characteristics of these events with respect to pure nuclear recoils,
which may provide additional suppression. In addition, a narrow conducting ring around the outer
circumference of the active volume just below the gas layer could be
used to intercept and absorb
the electrons drifting upwards from surface events, thus preventing
these events from having  the secondary burst of electro-luminescence light and
causing them to be rejected from the analysis. DarkSide is currently
investigating all of these rejection techniques in order to determine
the combination which mitigates the surface background with the least
impact on the fiducial volume.

Finally, electron recoil backgrounds in DarkSide will be suppressed using the
pulse shape discrimination and charge-to-energy based techniques described
earlier, at levels in excess of 10$^{-10}$. 

Detailed Monte Carlo simulations of the DarkSide-50 detector
conservatively suggest that the residual background rate in the WIMP
recoil energy region in DarkSide-50 will be $<$0.05
ev/0.1\,T-yr ($<$0.1 ev/0.1\,T-yr using the R11065 PMTs) after the application of the active
suppression techniques described above. As the (small) number of each
class of background events removed by the active veto techniques will
be known, and the efficiencies of the veto techniques themselves
calibrated\footnote{The efficiency of the neutron veto and the
  electron recoil and surface background discrimination can be
  calibrated directly using appropriate radioactive sources; the
  veto efficiency for cosmogenic backgrounds can be understood by
  comparing the number of cosmogenic events seen in the water tank, the
neutron veto and/or the argon detector with Monte Carlo expectation.},
DarkSide should be able to make a convincing claim that this very small
residual background is reasonable.

\section{DarkSide-10}
\label{sec:ds10}

\begin{figure}[htb]
\centering
\begin{minipage}[c]{7cm}
  \includegraphics[height=.25\textheight]{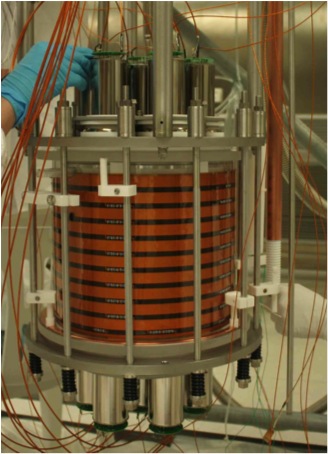}
\end{minipage}
\begin{minipage}[c]{7cm}
  \includegraphics[height=.25\textheight]{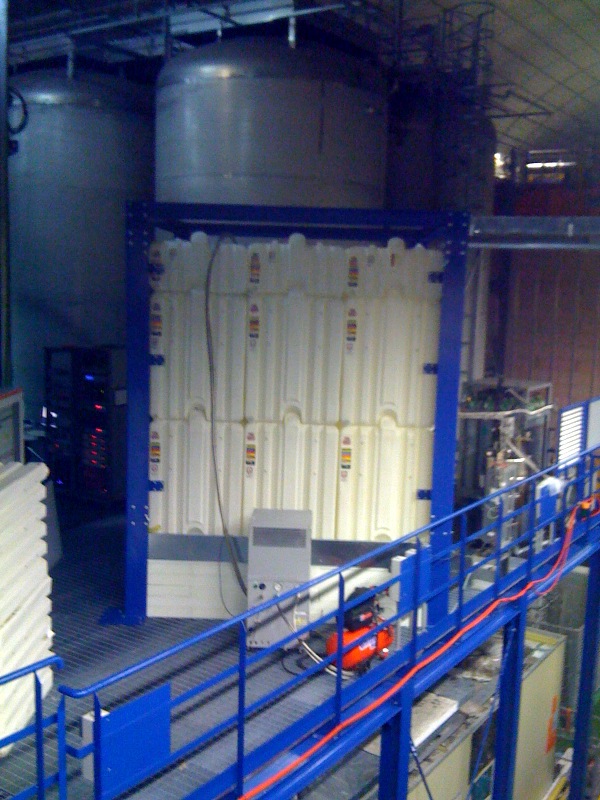}
\end{minipage}
\caption{Left: The DarkSide-10 inner detector as it was prepared for
  deployment at LNGS. Right: The passive shielding structure in which
  DarkSide-10 is currently (August 2011) being operated in Hall C of LNGS.}
\label{fig:DS10}
\end{figure}

A 10\,kg prototype detector, shown in Figure \ref{fig:DS10}, has been
built to test some key DarkSide technologies. The detector was
operated at Princeton University in two campaigns totaling about 7
months of running time during late 2010 and early 2011.

The Princeton runs demonstrated several important aspects of the
DarkSide design:
\begin{enumerate}
\item Transparent Indium-Tin-Oxide coatings on the top and bottom
  windows of the detector were successfully used to provide uniform
  anode and cathode planes. These are expected to provide a more uniform
  electric field in the gas multiplication region than would wire grids,
  and they also eliminate the ``dead'' gas regions between grid layers which can distort background
  events.
\item The gas pocket necessary for the production of
  electro-luminescence light was successfully created and stably
  maintained beneath the
  liquid argon which covers the upper photodetectors in the DarkSide design.
\item The R11065 PMTs performed to specifications during extended
  running.
\item The liquid argon was successfully purified to allow electron
  drift lifetimes of $>$300\,$\mu$s, sufficient for DarkSide-50.
\item A reasonably high light yield, $\sim$4.5 p.e./keV, was
obtained in the Princeton deployments. This is less than the 6 p.e./keV
that is targeted for DarkSide-50; however, given a number of {\it a
  priori} known optical non-idealities in the DarkSide-10
configuration in the Princeton deployments, we believe that DarkSide-50 will comfortably achieve 6 p.e./keV.
\item Sustained two-phase operation was successfully achieved,
  allowing data acquisition and analysis procedures that will be used
  on DarkSide-50 to be tested and improved. Although the TPC was
  operated with electric
  fields below its design value (due to an unstable high
  voltage feed-through), electron discrimination ability was seen in
  both the PSD and charge/light channels.
\end{enumerate}

In order to more carefully study background rejection in DarkSide-10,
a lower background environment was needed. To this end, the detector
has been moved to LNGS, where it has been installed in Hall C within a passive
shield consisting of $>$24'' of water/polyethylene shielding, which is
also shown in Figure \ref{fig:DS10}. Some upgrades were made during
the transfer process, which we hope will result in an increased light
yield. The malfunctioning high voltage feed-through was also
replaced. The detector has been re-filled with liquid argon, and is
currently being re-commissioned. Once this is complete we plan
to undertake detailed studies of electron recoil discrimination and
surface background rejection in the DarkSide detector configuration.

\section{The Future DarkSide Program}
\label{sec:dsFut}

\begin{figure}[htb]
\centering
\includegraphics[height=.3\textheight]{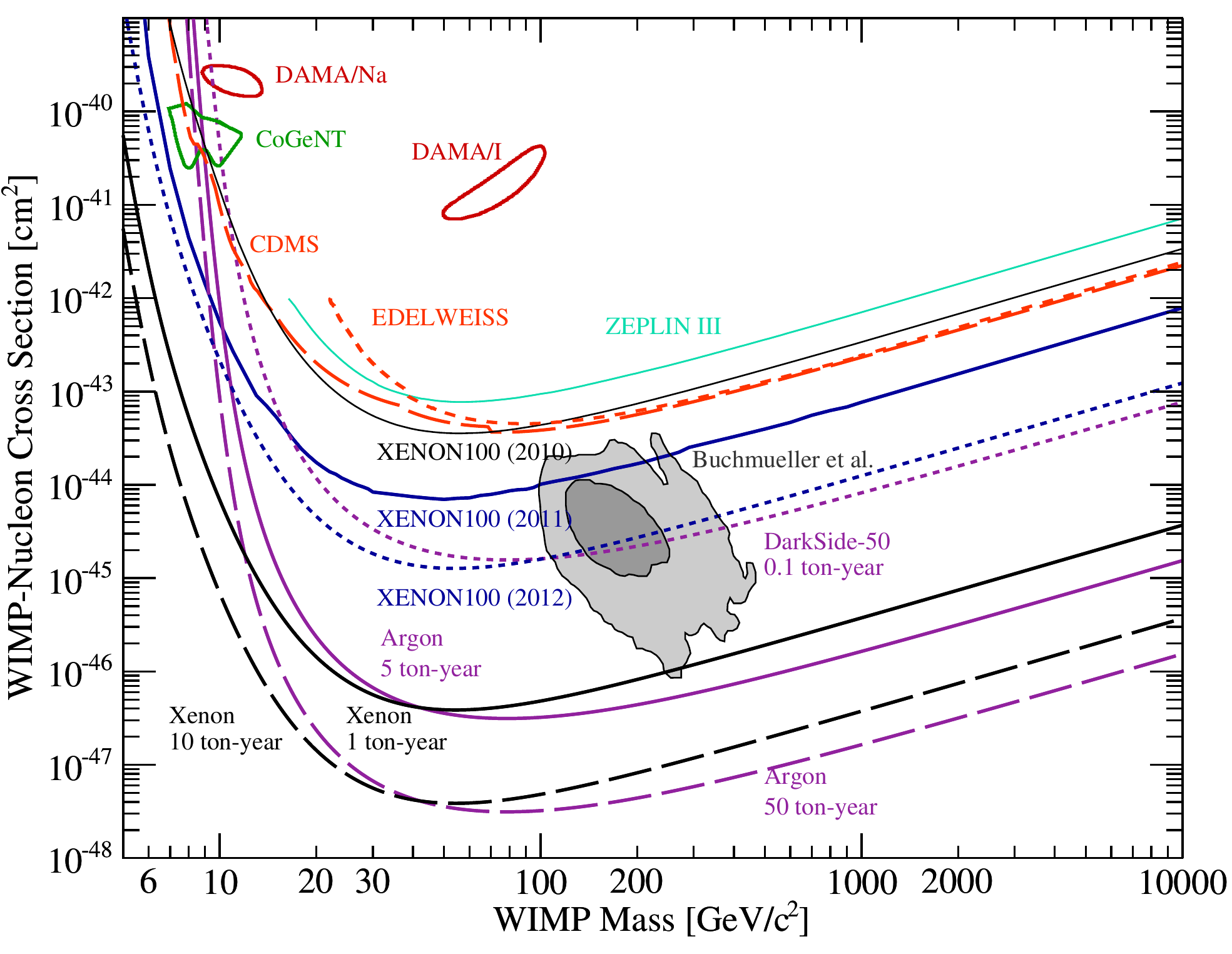}
\caption{Projected sensitivities of different detectors in the
  DarkSide/MAX program. Light yields of 6 p.e./keV$_{ee}$, and an
  $^{39}$Ar activity of 40\,mBq/kg (a depletion factor of 25 compared
  to atmospheric argon) were assumed in making this Figure. In DarkSide-50, this imposes an analysis
  threshold of 30\,keV$_{\text{recoil}}$ to ensure that the PSD (the
  efficiency of which depends strongly on the total number of detected
  photoelectrons) is sufficient to suppress this electron recoil
  background - in the larger detectors the analysis threshold is
  higher. A 50\% PSD nuclear recoil acceptance is assumed. Depending on the true $^{39}$Ar activity in the depleted
  argon and the actual light output achieved, lower energy
  thresholds may be possible. This would
  result in improved sensitivities relative to the plotted curves,
  especially at lower WIMP mass.}
\label{fig:sensitivity}
\end{figure}

The first physics detector in the DarkSide program will be DarkSide-50. This 50\,kg detector will be constructed
using radiopure materials and deployed within the active shielding
described in Section \ref{sec:bg}. The DarkSide-50 design is quite mature,
and the materials screening program to identify materials of
sufficient radiopurity for use in the detector is well underway. The
contract for the construction of the stainless steel sphere that will
contain the liquid scintillator neutron veto has been
awarded, and plans are being made for the conversion of the
Borexino CTF tank. DarkSide-50 is scheduled to be deployed in late
2012, after which it should achieve a WIMP sensitivity of at least
10$^{-45}$cm$^2$ (see Figure \ref{fig:sensitivity}) in about three
years of operation. Thanks to the very low and well understood
backgrounds afforded by the low background design and active
shielding, the experiment could make a credible claim of dark matter
detection at similar cross section.  

The neutron veto for DarkSide-50 will be 2\,m in radius, which is
large enough to house a future 5 tonne detector. Other elements of the
DarkSide-50 infrastructure, for example the gas handling system, will
also be built with enough spare capacity to facilitate this
upgrade. As shown in Figure \ref{fig:sensitivity}, a 5\,T detector
could achieve a sensitivity of about 10$^{-47}$cm$^2$.

Finally, the technology developed by DarkSide could become part of the
MAX (``Multi-tonne Argon and Xenon'') experiment, which is envisioned
to consist of twin multi-tonne argon and xenon detectors. This large
experiment would probe dark matter interactions to the
``ultimate'' 10$^{-48}$cm$^2$ level, below which the irreducible
background from coherent neutrino-nucleus scattering dominates.

\section{Conclusions}

DarkSide is a vigorous research program aimed at directly detecting
WIMP-type dark matter with two-phase argon time projection
chambers. The collaboration is currently operating a 10\,kg prototype detector in
Hall C of LNGS, and is in the process of constructing DarkSide-50, a
50\,kg physics detector. DarkSide-50 should be deployed in late 2012,
with future upgrades to tonne-scale detectors and beyond possible. A combination of innovative low background techniques,
including naturally depleted argon and ultra-low background
photodetectors and active background suppression, will give the
DarkSide detectors extremely low levels of background. The use
of active background suppression will allow the background levels to
be measured {\it in situ}, which would greatly enhance the credibility
of any potential claim of dark matter detection. We believe that
this careful approach to the understanding of our backgrounds
should enable the DarkSide program to make
important contributions as the field pushes to ever more sensitive
experiments. 

\begin{acknowledgments}
The DarkSide program is supported in the USA by the NSF and the
DoE. We gratefully acknowledge the hospitality of Gran Sasso National
Laboratory (LNGS). The
author acknowledges the support of the Princeton University PFEP program. 
\end{acknowledgments}

\bigskip 

\bibliography{bibliography}




\end{document}